\documentclass[conference]{IEEEtran}
\IEEEoverridecommandlockouts
\usepackage{cite}
\usepackage{amsmath,amssymb,amsfonts}
\usepackage{booktabs}
\usepackage{algorithm}
\usepackage{algorithmic}
\usepackage{graphicx}
\usepackage{subfigure}
\usepackage{cleveref} 
\usepackage{textcomp}
\usepackage{xcolor}
\usepackage{geometry}
\def\BibTeX{{\rm B\kern-.05em{\sc i\kern-.025em b}\kern-.08em
    T\kern-.1667em\lower.7ex\hbox{E}\kern-.125emX}}
\begin{document}

\title{Joint Beamforming Design and Satellite Selection for Integrated Communication and Navigation in LEO Satellite Networks\\


}

\author{\IEEEauthorblockN{Jiajing Li, Binghong Liu, and Mugen Peng}
\IEEEauthorblockA{State Key Laboratory of Networking and Switching Technology, \\
Beijing University of Posts and Telecommunications,
Beijing, 100876, China. \\
Email: $\{$lijiajing, binghong$\_$liu, pmg$\}$@bupt.edu.cn}
}

\maketitle

\begin{abstract}
Relying on the powerful communication capabilities and rapidly changing geometric configuration, the Low Earth Orbit (LEO) satellites have the potential to offer integrated communication and navigation (ICAN) services. However, the isolated resource utilization in the traditional satellite communication and navigation systems has led to a compromised system performance. Against this backdrop, this paper formulates a joint beamforming design and satellite selection optimization problem for the LEO-ICAN network to maximize the sum rate, while simultaneously reconciling the positioning performance. A two-layer algorithm is proposed, where the beamforming design in the inner layer is solved by the difference-of-convex programming method to maximize the sum rate, and the satellite selection in the outer layer is modeled as a coalition formation game to simultaneously reconcile the positioning performance. Simulation results verify the superiority of our proposed algorithms by increasing the sum rate by $16.6\%$ and $29.3\%$ compared with the conventional beamforming and satellite selection schemes, respectively. 
\end{abstract}


\section{Introduction}

With the rapid development of space technology, the low earth orbit (LEO) mega-constellations such as Starlink and OneWeb have experienced unprecedented growth, which not only provides abundant payloads, links, and terminal resources but also offers powerful communication coverage and information transmission capabilities\cite{b1,b2}. Meanwhile, the rapidly changing geometric configuration and strong power advantages of LEO satellites can effectively supplement, back up, and enhance the existing navigation systems\cite{b3}. Therefore, the LEO satellites have the potential to provide integrated communication and navigation (ICAN) services with wider coverage and higher precision\cite{b4}. Note that there is an interaction between communication and navigation. On the one hand, reliable communication transmission can enhance the receiver’s anti-interference performance and reduce error rate, thereby improving positioning accuracy; On the other hand, the navigation information can assist link management\cite{b5,frequency_precompensation} and resource optimization, thereby improving communication efficiency and reliability. However, the existing communication and navigation systems are mainly designed independently, which not only leads to the inefficient utilization of resources but also restricts the system performance improvement brought by the interaction between communication and navigation. Therefore, there is an urgent need to design and optimize the integrated communication and navigation system for LEO satellites (LEO-ICAN)  to achieve dual functions in communication and navigation.

Note that communication and navigation have distinct requirements for signals. The communication emphasizes information transmission, and its performance evaluation indices include rate and capacity. The navigation emphasizes the positioning calculation, and its performance evaluation indices include accuracy and Geometric Dilution of Precision (GDOP)\cite{b6}. Extensive researches have been conducted on resource scheduling in LEO-ICAN systems to enhance communication or navigation performance. For instance, in \cite{b7}, the authors investigated the beamforming design in distributed antenna systems to minimize power consumption. In \cite{b8}, the authors focused on the NOMA-based LEO satellite system and jointly optimized the power allocation and beam pattern selection to minimize the capacity-demand gap. In \cite{b9}, the authors considered cooperative positioning in the LEO satellite network and jointly optimized the beam scheduling and beamforming design to improve positioning accuracy. In \cite{b10}, the authors derived the expression of weighted GDOP and proposed a GDOP-based algorithm to optimize the geometric distribution of base stations for positioning. 

Most of the previous works only focused on the single performance index\cite{b8,b9,b10}, which leads to the difficulty of balancing communication and navigation performance. In addition, there is a tradeoff between the throughput and GDOP, for instance, more LEO satellites benefit the network topology but make the interference more complicated as well.
Therefore, in this paper, we take both the communication throughput and the navigation GDOP into consideration, explore their relationship, and further reveal their dependence on beamforming design and satellite selection. The main contributions of this work are summarized as follows:
\begin{itemize}
    \item We focus on the LEO-ICAN systems, where LEO satellites broadcast ICAN signals to the user equipments (UEs). We formulate the communication throughput maximization problem by jointly optimizing the beamforming design and satellite selection schemes, while simultaneously reconciling the GDOP performance for the navigation.
    \item A two-layer iterative algorithm is proposed to deal with the optimization problem. Specifically, in the inner layer, the Difference of Convex (DC) programming-based algorithm is proposed to obtain the beamforming design method. In the outer layer, the satellite selection problem is modeled as a coalition formation game, in which the data rate and GDOP are both taken into consideration to seek UEs' optimal satellite coalition.
    \item Simulation results demonstrate the superiority of our proposed algorithms. Compared with the conventional beamforming and satellite selection schemes, the sum rate can be improved by $16.6\%$ and $29.3\%$, respectively.
\end{itemize}
\addtolength{\topmargin}{0.05in}
\section{System Model and Problem Formulation}

\begin{figure}[htbp]
\centering
\includegraphics[width=0.45\textwidth]{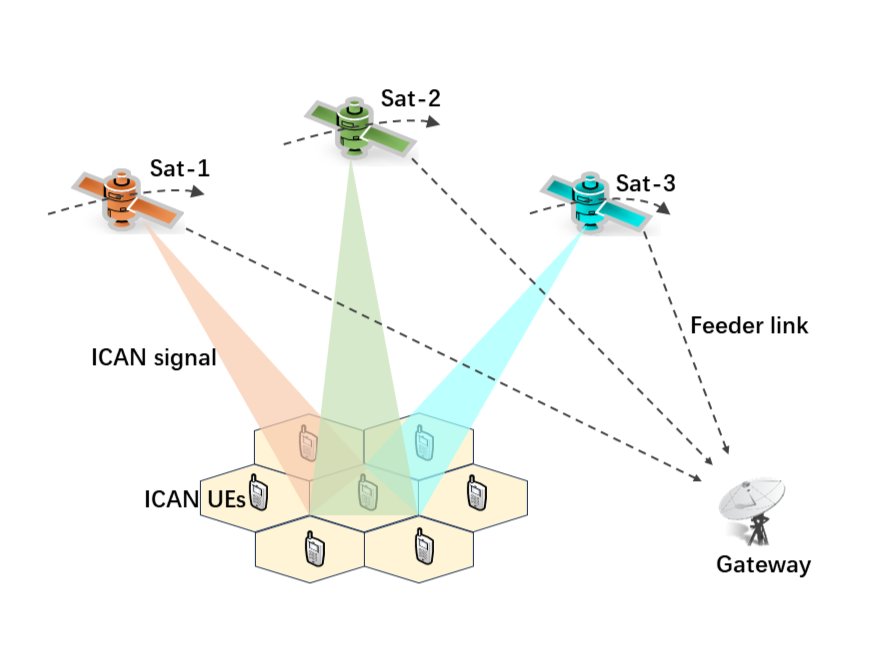}
\caption{Architecture of the LEO-ICAN system}
\label{fig:scenraio}       
\end{figure}

The LEO-ICAN system with $S$ LEO satellites and $C$ earth-fixed cells is illustrated in Fig. \ref{fig:scenraio}. The sets of satellites and cells are denoted as $\mathcal{S}=\left\{1,..., S\right\}$ and $\mathcal{C}=\left\{1,...,C\right\}$, respectively. For simplicity, it is assumed that each cell contains only one UE, thus UE $c$ represents the user equipment located in cell $c$. Each UE is equipped with a single antenna and each satellite is equipped with a uniform planar array (UPA), broadcasting the ICAN signals and providing communication and navigation services for UEs. The gateway stations are utilized to implement satellite selection and beamforming, and further forward the results to LEO satellites via feeder links. The binary variable $\alpha_{s,c}$ is introduced as a service indicator, where $\alpha_{s,c}=1$ means satellite $s$ serves UE $c$, and $\alpha_{s,c}=0$ otherwise. To achieve high-accuracy navigation, $I$ satellites are selected for each UE. $\mathcal{C}_{s}$ and $\mathcal{S}_{c}$ represent the set of UEs served by satellite $s$ and the set of satellites serving UE $c$, respectively.
\vspace{-1.0em}

\subsection{Channel Model}
As referred to 3GPP TR38.811\cite{b11}, the free path loss and the atmosphere attenuation are both taken into consideration to model the satellite-terrestrial channel, which is given by \cite{b12}
\begin{equation}
{{\bf{h}}_{s,c}} = \sqrt {G_{s,c}^{pl}G_{s,c}^{at}N} {{\rm{e}}^{ - j{\theta _{s,c}}}}{{\bf{v}}_{s,c}},
\end{equation}
where $G_{s,c}^{pl} = {\left( {\frac{\lambda }{{4\pi {d_{s,c}}}}} \right)^2}$ represents the free path loss with $\lambda=c/f$ representing the wavelength, and $d_{s,c}$ being the distance between satellite $s$ and UE $c$. $G_{s,c}^{at}$ represents the atmosphere attenuation. $N=N_{x}N_{y}$ is the number of UPA antennas, where $N_{x}$ and $N_{y}$ are the numbers of UPA antennas on the $x$ and $y$ axis, respectively. ${\theta _{s,c}}$ is the random phase vector with a uniform distribution in $[0,2\pi)$. ${{\bf{v}}_{s,c}}$ is the UPA response vector, which can be further expressed according to \cite{b13}
\begin{equation}
    {{\bf{v}}_{s,c}} = {{\bf{v}}_x} \otimes {{\bf{v}}_y},
\end{equation}
\begin{equation}
    {{\bf{v}}_x(\theta_{s,c}^x)} = \frac{1}{{\sqrt {{N_x}} }}{\left[ {1,{{\rm{e}}^{ - j\pi \theta_{s,c}^x}}, \ldots ,{{\rm{e}}^{ - j\pi \left( {{N_x} - 1} \right)\theta_{s,c}^x}}} \right]^T}, \notag
\end{equation}
\begin{equation}
     {{\bf{v}}_y(\theta_{s,c}^y)}= \frac{1}{{\sqrt {{N_y}} }}{\left[ {1,{{\rm{e}}^{ - j\pi \theta_{s,c}^y}}, \ldots ,{{\rm{e}}^{ - j\pi \left( {{N_y} - 1} \right)\theta_{s,c}^y}}} \right]^T}, \notag
\end{equation}
where ${{\bf{v}}_x}$ and ${{\bf{v}}_y}$ are the response vectors with respect to $x$ and $y$ axis, respectively. $\theta_{s,c}^x=\sin \varphi _{s,c}^y\cos \varphi _{s,c}^x$ and $\theta_{s,c}^y=\cos \varphi _{s,c}^y$ with $\varphi _{s,c}^y$ and $\varphi _{s,c}^y$ being the angles of the propagation path from satellite $s$ to UE $c$ with respect to the $x$ and $y$ axis, respectively. $\otimes$ represents the Kronecker product.

\subsection{Performance Metric}
To accurately assess the efficiency of the LEO-ICAN system, we consider the following two metrics from the communication and navigation aspects, respectively.
\subsubsection{Data Rate}
The ICAN signal received by UE $c$ from LEO satellite $s$ can be given by 
\begin{equation}
    {y_{s,c}} = {\bf{h}}_{s,c}^H{{\bf{w}}_{s,c}}{x_{s,c}} + \sum\limits_{c' \in {{\cal C}_s},c' \ne c} {{\bf{h}}_{s,c}^H{{\bf{w}}_{s,c'}}{x_{s,c'}}}  + {n_{s,c}},
\end{equation}
where ${{\bf{w}}_{s,c}}$ is the beamforming vector. $x_{s,c}$ is the ICAN symbol, which contains both communication and navigation data. $n_{s,c}\sim\mathcal{N}(0,\sigma _c^2)$ is the noise received by UE $c$.
The orthogonal frequencies are allocated among LEO satellites to mitigate the inter-satellite interference, while full frequency reuse among beams is applied within the same satellite, introducing the intra-satellite interference. As such, the signal-to-interference-plus-noise ratio (SINR) is given by
\begin{equation}
    {\rm{SIN}}{{\rm{R}}_{s,c}} = \frac{{{{\left| {{\bf{h}}_{s,c}^H{{\bf{w}}_{s,c}}} \right|}^2}}}{{\sum\limits_{c' \in {{\cal C}_s},c' \ne c} {{{\left| {{\bf{h}}_{s,c}^H{{\bf{w}}_{s,c'}}} \right|}^2}}  + \sigma _c^2}}.
\end{equation}
Then, the data rate of UE $c$ served by LEO satellite $s$ is given by
\begin{equation}
\label{eq: data rate}
    {R_{s,c}} = B{\log _2}(1 + {\rm{SIN}}{{\rm{R}}_{s,c}}),
\end{equation}
where $B$ is the bandwidth allocated to each satellite.
\subsubsection{GDOP}
Network topology, which is usually assessed by GDOP, plays a crucial role in the navigation service. 
We denote the three dimensional coordinates of UE $c$ and satellite $s$ by ${{\bf{p}}_c} = {[{x_c},{y_c},{z_c}]^T}$ and ${{\bf{p}}_s} = {[{x_s},{y_s},{z_s}]^T}$, respectively. The distance between satellite $s$ and UE $c$ is given by
\begin{equation}
    {d_{s,c}} = {\left\| {{{\bf{p}}_c} - {{\bf{p}}_s}} \right\|_2}.\
\end{equation}

To simplify the model, we assume that the clock error is fixed. According to \cite{b_gdop}, the GDOP value of UE $c$ can be calculated by
\begin{equation}
    {{\rm{GDOP}}_c} = \sqrt {\mathrm{tr}({{\left( {{G_c}^T{G_c}} \right)}^{ - 1}})} ,
\end{equation}
where the elements in matrix $G_c$ are the partial derivatives of the observed distance between UE $c$ and its associated LEO satellites. Consequently, $G_c$ can be further expressed as
\begin{equation}
    G_c = \left( {\begin{array}{*{20}{c}}
{{x_{{s_1},c}}}&{{y_{{s_1},c}}}&{{z_{{s_1},c}}}\\
{{x_{{s_2},c}}}&{{y_{{s_2},c}}}&{{z_{{s_2},c}}}\\
 \cdots & \cdots & \cdots \\
{{x_{\left| {{{\cal S}_c}} \right|,c}}}&{{y_{\left| {{{\cal S}_c}} \right|,c}}}&{{z_{\left| {{{\cal S}_c}} \right|,c}}}
\end{array}} \right),
\end{equation}
where ${x_{s,c}} = \frac{{{x_c} - {x_s}}}{{{d_{s,c}}}}$, ${y_{s,c}} = \frac{{{y_c} - {y_s}}}{{{d_{s,c}}}}$, ${z_{s,c}} = \frac{{{z_c} - {z_s}}}{{{d_{s,c}}}}$ for any $s \in {{\cal S}_c}$. 

\subsection{Problem Formulation}
In this paper, we aim to maximize the communication throughput of all the UEs by jointly optimizing the beamforming design scheme $\bf{w}$ and satellite selection scheme $\alpha$, while simultaneously satisfying the GDOP constraint. The formulated optimization problem is given by
\begin{align} \label{P0}
&\mathop {\max }\limits_{{\alpha _{s,c}},{{\bf{w}}_{s,c}}} \sum\limits_{s \in {\cal S},c \in {{\cal C}_s}} {{R_{s,c}}} \\
s.t. \quad&{\alpha _{s,c}} \in \left\{ {0,1} \right\},\forall s \in {\cal S},c \in {\cal C}, \tag{\ref{P0}{a}} \label{P0a}\\
     &\sum\limits_s {{\alpha _{s,c}}}  = I, \forall c \in {\cal C}, \tag{\ref{P0}{b}} \label{P0b}\\
     &{\rm{GDO}}{{\rm{P}}_c} \le \gamma , \forall c \in {\cal C},\tag{\ref{P0}{c}} \label{P0c}\\
     &\left\| {{{\bf{w}}_{s,c}}} \right\|_2^2 \leq P \cdot {\alpha _{s,c}}, \forall s \in {\cal S},c \in {\cal C}, \tag{\ref{P0}{d}} \label{P0d}
\end{align}
where constraint (\ref{P0a}) means $\alpha _{s,c}$ is binary. Constraints (\ref{P0b}) and (\ref{P0c}) are the topology-related constraints, where (\ref{P0b}) regulates that each UE should be served by $I$ LEO satellites, and (\ref{P0c}) restricts the GDOP of each UE. Constraint (\ref{P0d}) is the transmission power constraint, where $P$ is the power budget of each beam.

\section{A joint algorithm for beamforming design and satellite selection}
It is evident that the optimization problem in (\ref{P0}) is a mixed-integer non-convex problem (MINCP) due to the existence of integer variable $\alpha _{s,c}$ and the sum-log objective function\cite{b14}. To tackle this issue, we decouple the original problem into the separate beamforming design subproblem and the satellite selection subproblem and propose a two-layer algorithm to iteratively optimize these two subproblems until convergence. 

\subsection{Inner layer optimization: Beamforming Design}
In this subsection, we first deal with the beamforming design subproblem in the inner layer with the fixed satellite selection scheme. At this point, the original problem can be simplified as
\begin{align} \label{P1}
&\mathop {\max }\limits_{{{\bf{w}}_{s,c}}} \sum\limits_{s \in {\cal S},c \in {{\cal C}_s}} {{R_{s,c}}} \\
s.t. &\qquad(\ref{P0d}). \notag
\end{align}

As mentioned above, there is no inter-satellite interference, and the beamforming design of each satellite is independent. Therefore, the problem of maximizing the sum rate of all UEs can be simplified to maximizing the sum rate of each satellite, which is given by
\begin{align} \label{P1_per_sat}
&\mathop {\max }\limits_{{{\bf{w}}_{s,c}}} \sum\limits_{c \in {{\cal C}_s}} {{R_{s,c}}} \\
s.t. \quad&\left\| {{{\bf{w}}_{s,c}}} \right\|_2^2 \leq P , \forall c \in {{\cal C}_s}. \tag{\ref{P1_per_sat}{a}} \label{P1a}
\end{align}
Then, the auxiliary variable ${{\bf{Q}}_{s,c}} = {{\bf{w}}_{s,c}}{\bf{w}}_{s,c}^H$ is introduced to transform problem (\ref{P1_per_sat}) into the following form
\begin{align} \label{P2}
&\mathop {\max }\limits_{{{\bf{w}}_{s,c}}} \sum\limits_{c \in {{\cal C}_s}} {{R_{s,c}}} \\
s.t. \quad&{\rm{trace}}({{\bf{Q}}_{s,c}}) \leq P, \forall c \in {{\cal C}_s}, \tag{\ref{P2}{a}} \label{P2:power constraint}\\
&{\rm{rank}}({{\bf{Q}}_{s,c}}) = 1, \forall c \in {{\cal C}_s}, \tag{\ref{P2}{b}} \label{P2:rank-1 constraint}\\
&{{\bf{Q}}_{s,c}}\succeq0, \forall c \in {{\cal C}_s}. \tag{\ref{P2}{c}} \label{P2:PSD constraint}
\end{align}

Note that both the constraint (\ref{P2:rank-1 constraint}) and objective function are non-convex, making the problem hard to solve. To deal with the non-convex constraint (\ref{P2:rank-1 constraint}), we utilize the rank-1 characteristic to relax (\ref{P2:rank-1 constraint}) according to \cite{b15}. To deal with the non-convex objective function, the DC programming \cite{b16}, can be employed to approximately convert it into the convex form. 
Specifically, the data rate can be further represented as
\begin{equation}\label{eq:data rate with Q}
\begin{aligned}
    {R_{s,c}} = &B{\log _2}(\sigma _c^2 + \sum\limits_{c \in {{\cal C}_s}} {h_{s,c}^H{{\bf{Q}}_{s,c}}{h_{s,c}}} ) \\ &- B{\log _2}(\sigma _c^2 + \sum\limits_{c' \in {{\cal C}_s},c \ne c} {h_{s,c}^H{{\bf{Q}}_{s,c'}}{h_{s,c}}} )\\
    &=f\left( {{{\bf{Q}}_{s,c}}} \right)-g\left( {{{\bf{Q}}_{s,c}}} \right).
\end{aligned}
\end{equation}
Due to the non-convexity of the log function, $f\left( {{{\bf{Q}}_{s,c}}} \right)$ in (\ref{eq:data rate with Q}) is concave, while $-g\left( {{{\bf{Q}}_{s,c}}} \right)$ is convex. 
At this point, the first order Taylor expansion of $g\left( {{{\bf{Q}}_{s,c}}} \right)$ at ${{{\bf{Q}}_{s,c}}}={{{\bf{Q}}_{s,c}^n}}$ is given by 

\begin{equation} \label{eq:taylor expansion of g}
    \begin{aligned}
        &\overline g ({\bf{Q}}_{s,c}^{n + 1};{\bf{Q}}_{s,c}^n) = B{\log _2}(\sum\limits_{c' \in {{\cal C}_s},c' \ne c} {{\bf{h}}_{s,c}^H{\bf{Q}}_{s,c'}^n{{\bf{h}}_{s,c}}}
        + \sigma _c^2)\\ &\quad+ \frac{B{\sum\limits_{c' \in {{\cal C}_s},c' \ne c} {({\bf{h}}_{s,c}^H{\bf{Q}}_{s,c'}^{n + 1}{{\bf{h}}_{s,c}} - {\bf{h}}_{s,c}^H{\bf{Q}}_{s,c'}^n{{\bf{h}}_{s,c}})} }}{{\ln 2(\sum\limits_{c' \in {{\cal C}_s},c' \ne c} {{\bf{h}}_{s,c}^H{\bf{Q}}_{s,c'}^n{{\bf{h}}_{s,c}}}  + \sigma _c^2)}},
    \end{aligned}
\end{equation}
where ${{{\bf{Q}}_{s,c}^n}}$ is the optimal ${{{\bf{Q}}_{s,c}}}$ in the last iteration.
So far, problem (\ref{P2}) can be approximately given by
\begin{align} \label{P2_final}
&\mathop {\max }\limits_{{\bf{Q}}_{s,c}^{n + 1}} \sum\limits_{c \in {{\cal C}_s}} {F({\bf{Q}}_{s,c}^{n + 1})}\\
s.t. \quad&{\rm{trace}}({{\bf{Q}}_{s,c}^{n + 1}}) \leq P, \forall c \in {{\cal C}_s}, \tag{\ref{P2_final}{a}} \\
&{\bf{Q}}_{s,c}^{n + 1}\succeq0, \forall c \in {{\cal C}_s}. \tag{\ref{P2_final}{b}} \label{P2_final:PSD constraint}
\end{align}
where $F({\bf{Q}}_{s,c}^{n + 1}) = f\left( {{{\bf{Q}}_{s,c}^{n + 1}}} \right) - \overline g({\bf{Q}}_{s,c}^{n + 1};{\bf{Q}}_{s,c}^n)$.

The DC programming-based beamforming algorithm is outlined in Algorithm \ref{Algorithm 1: BA}. Firstly, the beamforming vectors are randomly initialized within the feasible region. Then, iterations are performed step by step to approach a local optimal solution. In each iteration, we solve problem (\ref{P2_final}) by the CVX tools and update the solution ${\bf{Q}}_{s,c}^{n + 1}$. Such an operation continues until convergence to get the optimal solution ${\bf{Q}}_{s,c}^{opt}$. By applying rank-1 approximation\cite{b15}, the beamforming vector ${{\bf{w}}_{s,c}^{opt}}$ can be obtained via ${{\bf{w}}_{s,c}^{opt}} = \sqrt {{\lambda _{\max }}} {{\bf{b}}_{\max }}$, where $\lambda _{\max }$ and ${\bf{b}}_{\max }$ are the largest eigenvalue and the corresponding eigenvector of ${\bf{Q}}_{s,c}^{opt}$, respectively.

\begin{algorithm}[htbp]

    \setcounter{algorithm}{0}
	\caption{DC programming-based Beamforming Algorithm }
    \label{Algorithm 1: BA}
    \begin{algorithmic}[1] 

        \FORALL {satellite $s=1,2,\cdots,S$}
                \STATE Randomly initialize ${\bf{Q}}_{s,c}^1$ within the feasible region.
                \STATE $n=1$.
            \WHILE {$\sum\limits_{c \in {{\cal C}_s}} {\left| {F({\bf{Q}}_{s,c}^{n + 1}) - F({\bf{Q}}_{s,c}^n)} \right|}  \ge \delta $} 
                \STATE Solve the optimization problem (\ref{P2_final}).
                \STATE Update ${\bf{Q}}_{s,c}^{n+1}$.
                \STATE $n=n+1$.
            \ENDWHILE
            \STATE \textbf{return} ${\bf{Q}}_{s,c}^{opt}$.
            \STATE Apply rank-1 approximation of ${\bf{Q}}_{s,c}^{opt}$ to get ${\bf{w}}_{s,c}^{opt}$.
        \ENDFOR
    \end{algorithmic}
\end{algorithm}

\subsection{Outer layer optimization: Satellite Selection}
In this section, we deal with the satellite selection subproblem in the outer layer with the fixed beamforming design scheme. At this point, the original optimization problem can be simplified as
\begin{align} \label{P3}
&\mathop {\max }\limits_{{\alpha _{s,c}}} \sum\limits_{s \in {\cal S},c \in {{\cal C}_s}} {{R_{s,c}}} \\
s.t. &\quad(\ref{P0a}), (\ref{P0b}), (\ref{P0c}). \notag
\end{align}

Note that the optimization problem (\ref{P3}) is an integer programming problem. To obtain an effective solution, we model the satellite selection problem as a coalition formation game (CFG)\cite{b17}, in which each UE forms a coalition with $I$ satellites.

Specifically, the coalition set of all UEs is denoted by ${\cal M} = \left\{ {{{\cal M}_c}\left| {c \in {\cal C}} \right.} \right\}$, where ${{\cal M}_c}$ represents the coalition of UE $c$ and ${{\cal M}_c} \subset {\cal S}$. If there exists satellite $s \in {{\cal M}_c}$, then UE $c$ is served by satellite $s$, namely ${\alpha _{s,c}} = 1$. We utilize the sum rate of all UEs as the utility function of the coalition game, which is given by
\begin{equation}
    U = \sum\limits_{c \in {\cal C},s \in {{\cal S}_c}} {{R_{s,c}}}.
\end{equation}

A feasible coalition is required to satisfy the following properties:

\begin{enumerate}
\item $\left| {{{\cal M}_c}} \right| = I, \forall c \in {\cal C}$ \label{property 1};
\item $GDO{P_c}({\cal M}) \le \gamma, \forall c \in {\cal C} $\label{property 2}.
\end{enumerate}
The property \ref{property 1}) is related to the constraint (\ref{P0b}) and indicates that each UE should be served by $I$ satellites, the property \ref{property 2}) is related to the constraint (\ref{P0c}) and regulates the GDOP value that needs to be met. 

The CFG-based satellite selection algorithm is given in Algorithm 2, which mainly consists of the initialization phase and the coalition switch phase. In the initialization phase, the GDOP-based satellite selection is utilized to establish the initial coalition. In the coalition switch phase, each UE attempts to switch to new coalitions for a higher utility. The coalition switching rule is guided by UEs' preference lists, which are arranged in ascending order according to the feasible GDOP values. Eventually, the optimal coalition can be determined when further coalition switches cannot lead to a higher utility. Compared to random switching, the proposed method can not only ensure the GDOP constraint but also maximize the sum rate with fewer switch times.

\begin{algorithm}[htbp]
    \setcounter{algorithm}{1}
	\caption{CFG-based Satellite Selection Algorithm}
    \label{Algorithm 2: SSA}
    \begin{algorithmic}[1] 
     \STATE $\textbf{Initialization:}$
        \STATE Initialize the coalition structure ${{\cal M}^{old}}$ according to the GDOP-based satellite selection algorithm.
        \STATE Obtain the initial beamforming vector ${\bf{w}}_{s,c}^{old}$ and coalition utility $U^{old}$ according to Algorithm \ref{Algorithm 1: BA}.
        \STATE $\textbf{Coalition switch:}$
        \FORALL {UE $c=1,2,\cdots,C$}
            \STATE Construct its preference list ${{\cal R}_{c}}$ according to the feasible GDOP values.
            \FORALL {satellite combinations in ${{\cal R}_{c}}$}
                    \STATE UE $c$ selects the current satellite combination to receive services.
                    \STATE Optimize the beamforming vector ${\bf{w}}_{s,c}^{new}$ according to Algorithm \ref{Algorithm 1: BA}.
                    \STATE Calculate the current coalition utility $U^{new}$.
                    \IF {$U^{new} \ge U^{old}$}
                        \STATE UE $c$ executes a coalition switch.
                        \STATE Update ${{\cal M}^{old}}$, ${\bf{w}}_{s,c}^{old}$ and $U^{old}$.
                    \ELSE
                        \STATE ${{\cal M}^{old}}$,  ${\bf{w}}_{s,c}^{old}$ and $U^{old}$ remain the same.
                    \ENDIF
            \ENDFOR
        \ENDFOR
        \STATE The optimal satellite selection scheme $\alpha _{s,c}$ decided by the final ${{\cal M}^{old}}$ and the corresponding beamforming vector ${\bf{w}}_{s,c}$ are taken as the solution to problem (\ref{P0}).
    \end{algorithmic}
\end{algorithm}

\section{Simulation}
In the simulation part, we consider a LEO-ICAN system with 7 satellites and 7 cells. Each UE is served by 3, 4, or 5 satellites. Other parameters utilized throughout the whole simulation are summarized in Table \ref{tabel: Simulation parameter} unless otherwise specified. 
\begin{table}[htbp]
\centering
\caption{Simulation parameter settings}
\label{tabel: Simulation parameter}
\begin{tabular}{c c}

\toprule
\textbf{Parameter}   & \textbf{value}   \\
\midrule
 The height of satellites & 600 km \\
 The radius of cells & 43.3 km \\
 The number of antennas per satellite & 8$\times$8 \\
 Operation frequency & 4 GHz \\
 Bandwidth & 50 MHz \\
 Maximum beam transmission power  & 26 dBw \\
 Noise power density & -174 dBm/Hz \\
 GDOP threshold $\gamma$  & 6 \\
 Beamforming convergence threshold $\delta$  & 0.5 Mbps \\
 \bottomrule
\end{tabular}
\vspace{-1.0em}
\end{table}

We introduce three comparison schemes to verify the superiority of our proposed algorithm, including two beamforming algorithms and one satellite selection algorithm, which are respectively given by

\begin{itemize}
    \item Single-cell beamforming algorithm\cite{b18}: in which the beamforming vector is given by
\begin{equation}\label{eq: SCB}
    \mathbf{w}_{s,c}=\sqrt{\frac P{\|\mathbf{h}_{s,c}\|_F^2}}\mathbf{h}_{s,c}.
\end{equation}
\item Zero-forcing (ZF) beamforming algorithm\cite{b19}: in which the intra-satellite interference caused by the frequency reuse is eliminated. Define $\mathbf{H}_{s} = [\mathbf{h}_{s,c_{1}},\cdots,\mathbf{h}_{s,c_{|\mathcal{C}_{s}|}}]^{H} \in \mathbb{C}^{|\mathcal{C}_{s}|\times N}$ as the channel vector of satellite $s$ and $\mathbf{H}_s^\dagger=\mathbf{H}_{s}^{H}(\mathbf{H}_{s}\mathbf{H}_{s}^{H})^{-1}$ as the pseudo-inverse of $\mathbf{H}_s$, respectively. Then, the beamforming matrix of satellite $s$ is given by 
\begin{equation}
    \mathbf{W}_s=\beta\mathbf{H}_s^\dagger
\end{equation}
with 
\begin{equation}
    \beta=\sqrt{\frac{P|\mathcal{C}_s|}{\|\mathbf{H}_s^\dagger\|_F^2}}. \notag
\end{equation}
\item GDOP-based satellite selection algorithm\cite{b20}: in which each UE selects satellites based on the principle of minimizing GDOP.
\end{itemize} 

For comparison, we also provide the following three baselines:
\begin{itemize}
    \item Baseline 1: The satellite selection and beamforming are optimized by the CFG-based and single-cell beamforming algorithms, respectively.
    \item Baseline 2: The satellite selection and beamforming are optimized by the CFG-based and ZF beamforming algorithms, respectively.
    \item Baseline 3: The satellite selection and beamforming are optimized by the GDOP-based and DC-based beamforming algorithms, respectively.
    \item Our proposal: The satellite selection and beamforming are optimized by the CFG-based and DC-based beamforming algorithms, respectively.
\end{itemize}
\vspace{-1.0em}
\begin{table}[htbp]
\centering
\caption{Sum rate under different schemes}
\label{tabel: sum rate under different schemes}
\begin{tabular}{c c c c}
\toprule
\textbf{Scheme}   & \textbf{Sum rate (Gbps)}   &\textbf{Scheme}   & \textbf{Sum rate (Gbps)}    \\
\midrule
Baseline 1 & 1.5852 & Baseline 3 & 2.4060\\
Baseline 2 & 2.6688 & Our proposal & 3.1110\\
 \bottomrule
\end{tabular}
\end{table}

Table \ref{tabel: sum rate under different schemes} demonstrates the sum rate under different beamforming and satellite selection schemes with $I=4$. It is observed that our proposal outperforms the ZF beamforming algorithm and the GDOP-based satellite selection algorithm by $16.6\%$ and $29.3\%$, respectively, achieving the best performance in maximizing the sum rate. This is due to the fact that the single-cell beamforming algorithm only focuses on the channel conditions between UEs and their corresponding service satellites without considering the interference. The ZF beamforming algorithm, while effectively eliminating the intra-satellite interference, inevitably amplifies the additive noise, thereby reducing the rate.  Meanwhile, compared with the GDOP-based satellite selection algorithm, our proposal effectively restrains the inter-cell interference by appropriately selecting satellites.
\vspace{-0.5em}
\begin{figure}[htbp]
\centering
\includegraphics[width=0.36\textwidth]{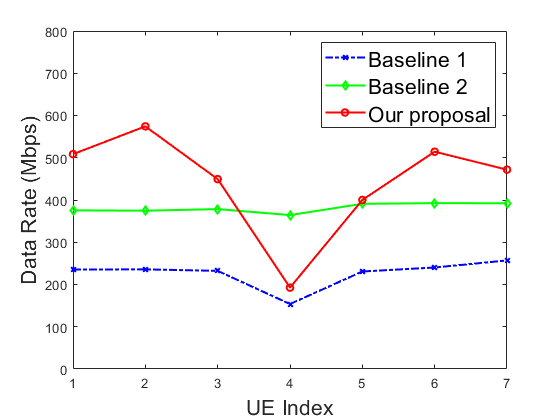}
\caption{Sum rate under different beamforming schemes.}
\label{fig: BA}
\end{figure}

In Fig. \ref{fig: BA}, the data rate of each UE under different beamforming algorithms is depicted. By applying the DC-based beamforming algorithm, the data rates of most UEs have experienced a significant increase compared with Baseline 1 and Baseline 2. Besides, it can be observed that the rate of UE 4 in our proposal is lower than that in Baseline 2. The reason is that UE 4 located in the central cell experiences and causes greater interference relative to other UEs. As such, the rate of UE 4 is moderately compromised to achieve the maximization of system throughput.
\vspace{-0.5em}
\begin{figure}[htbp]
	\centering
	\subfigure[GDOP value] {\includegraphics[width=.24\textwidth]{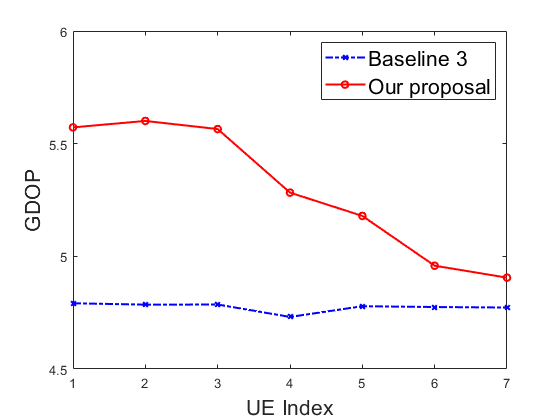}}
	\subfigure[Data rate] {\includegraphics[width=.24\textwidth]{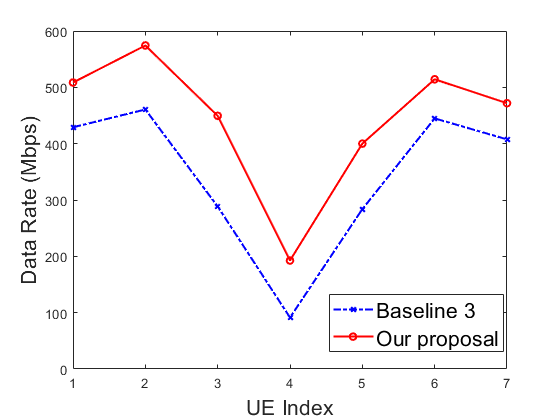}}
	\caption{Performance under different satellite selection schemes.}
	\label{fig: SSA}
\end{figure}

For Fig. \ref{fig: SSA}, the GDOP and rate performance of our proposed CFG-based satellite selection algorithm are presented. Note that in the traditional GDOP-based satellite selection algorithm, UEs tend to select similar satellite combinations for achieving the optimal topology, resulting in serious intra-satellite interference for some satellites, while leaving the other satellites underutilized. However, this issue has been effectively addressed within our proposed satellite selection algorithm. By appropriately relaxing the GDOP requirement, the proposed algorithm significantly improves the communication rate for UEs, achieving a balance between communication and navigation performance. 
\begin{figure}[htbp]
\centering
\includegraphics[width=0.36\textwidth]{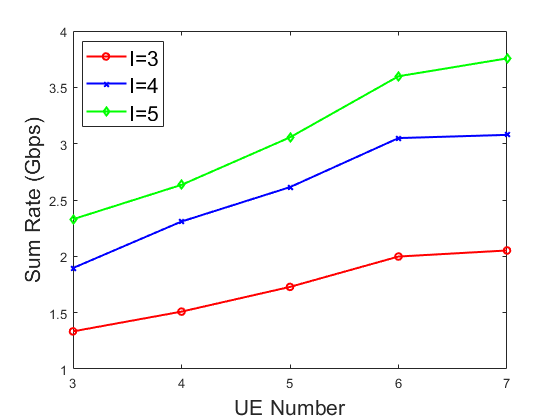}
\caption{Sum rate under different UE numbers.}
\label{fig: UE number}
\vspace{-1.0em}
\end{figure}

Fig. \ref{fig: UE number} illustrates the impact of UE number on the sum rate under the case of $I=3,4,5$. It is noteworthy that the sum rate increases monotonically with the number of UEs. However, this increase becomes flat due to the severe inter-cell interference caused by more UEs. Furthermore, the increase of $I$ not only enhances the GDOP performance but also improves the sum rate by serving each UE with more LEO satellites.

\section{Conclusion}
In this paper, joint beamforming design and satellite selection in the LEO-ICAN system have been investigated. To provide high-quality communication and navigation services, we formulate a throughput maximization problem, while simultaneously reconciling the GDOP performance. A two-layer algorithm is proposed, where the inner beamforming design is solved by the DC programming method, and the outer satellite selection is modeled as a coalition formation game. Through numerical results, we validate the effectiveness of the proposed algorithms compared to the existing benchmark schemes.

\section*{Acknowledgment}

This work was supported 
in part by the National Key Research and Development Program of China under Grant 2021YFB2900200,
in part by the National Natural Science Foundation of China under Grants 62401073 and 61925101,
in part by the Postdoctoral Fellowship Program of China Postdoctoral Science Foundation (CPSF) under Grant GZB20240081,
and in part by the Beijing Natural Science Foundation under Grant L223007.


\end{document}